\input epsf
\magnification \magstep1
\raggedbottom
\openup 1\jot
\voffset6truemm
\def\cstok#1{\leavevmode\thinspace\hbox{\vrule\vtop{\vbox{\hrule\kern1pt
\hbox{\vphantom{\tt/}\thinspace{\tt#1}\thinspace}}
\kern1pt\hrule}\vrule}\thinspace}
\def\II{{\rm1\!\hskip-1pt I}}
\centerline {\bf HIGHLIGHTS OF SYMMETRY GROUPS}
\vskip 1cm
\noindent
Giampiero Esposito and Giuseppe Marmo 
\vskip 0.3cm
\noindent
{\it INFN, Sezione di Napoli, and Dipartimento di Scienze Fisiche,
Complesso Universitario di Monte S. Angelo, Via Cintia, 
Edificio N', 80126 Napoli, Italy}
\vskip 1cm
\noindent
{\bf Abstract}. The concepts of symmetry and symmetry groups are
at the heart of several developments in modern theoretical and
mathematical physics. The present paper is devoted to a number of
selected topics within this framework: Euclidean and rotation 
groups; the properties of fullerenes in physical chemistry; Galilei,
Lorentz and Poincar\'e groups; conformal transformations and
the Laplace equation; quantum groups and Sklyanin algebras.
For example, graphite can be vaporized by laser irradiation,
producing a remarkably stable cluster consisting of $60$ carbon
atoms. The corresponding theoretical model considers a truncated
icosahedron, i.e. a polygon with $60$ vertices and $32$ faces, 
$12$ of which are pentagonal and $20$ hexagonal. The 
$C_{60}$ molecule obtained when a carbon atom is placed at each
vertex of this structure has all valences satisfied by two single
bonds and one double bond. In other words, a structure in which a
pentagon is completely surrounded by hexagons is stable. Thus,
a ``cage'' in which all $12$ pentagons are completely surrounded
by hexagons has optimum stability. On a more formal side, the 
exactly solvable models of quantum and statistical physics can
be studied with the help of the quantum inverse problem method.
The problem of enumerating the discrete quantum systems which can
be solved by the quantum inverse problem method reduces to the
problem of enumerating the operator-valued functions that satisfy 
an equation involving a fixed solution of the quantum 
Yang--Baxter equation. Two basic equations exist which provide a
systematic procedure for obtaining completely integrable lattice
approximations to various continuous completely integrable 
systems. This analysis leads in turn to the discovery of
Sklyanin algebras.
\vskip 100cm
\centerline {\bf 1.1 Introduction to symmetry groups}
\vskip 0.3cm
\noindent
Symmetry pervades all our descriptions, modelling or understanding
of natural phenomena. From microscopic physics to cosmology, from
chemistry to biology, symmetry properties or considerations are
ubiquitous in scientific research. 

In ancient times, symmetry meant mostly ``harmony in the proportions'',
and in modern natural science it emerges mostly as the ``invariance
with respect to a transformation group''. In abstract terms, any 
set $\cal S$ is associated with a group ${\rm Aut} \; {\cal S}$, the
family of all invertible maps $\varphi: {\cal S} \rightarrow {\cal S}$,
which may be composed:
$$
(\varphi_{1} \cdot \varphi_{2})(s)=\varphi_{1}(\varphi_{2}(s)),
\eqno (1.1.1)
$$
and the composition is associative, i.e.
$$
(\varphi_{1} \cdot \varphi_{2})\cdot \varphi_{3}
=\varphi_{1} \cdot (\varphi_{2} \cdot \varphi_{3}),
\eqno (1.1.2)
$$
it allows for the identity transformation
$$
\II: s \rightarrow s, \; \forall s \in {\cal S},
\eqno (1.1.3)
$$
and an inverse $\varphi^{-1}$ such that
$$
\varphi(\varphi^{-1}(s))=\varphi^{-1}(\varphi(s))
=\II \cdot s =s.
\eqno (1.1.4)
$$
Any subset, pattern, configuration $\cal P$ in $\cal S$ defines a
subgroup of ${\rm Aut} \; {\cal S}$; by selecting only those
transformations which preserve or keep invariant $\cal P$,
for any $p \in {\cal P} \subset {\cal S}$ we consider only those
transformations $\varphi \in {\rm Aut} \; {\cal S}$ such that
$\varphi(p) \in {\cal P}, \; \forall p \in {\cal P}$. This 
subgroup is never empty because it contains at least the identity
transformation which constitutes the smallest subgroup in
${\rm Aut} \; {\cal S}$. In addition to subsets, one may select
subgroups of ${\rm Aut} \; {\cal S}$, by requiring that properties
or relations among pairs or more points are preserved. This strict
connection between configurations and transformations was explicitly
introduced by F. Klein and S. Lie. In the Erlangen program (1872) of
F. Klein, a geometric theory is defined as the study of those 
properties of the space, $\cal S$, and of its subsets (figures) which are
preserved with respect to a selected subgroup of 
${\rm Aut} \; {\cal S}$. Similarly, in physics, starting with Galileo
and culminating with Einstein, every physical theory carries with it
its ``covariance'' group. 

It was P. Curie who (translated and) incorporated into physics the
role of symmetry as a working tool in our formalization of the external
world, by stating that ``symmetries in the cause should be reflected
in the effects, and the lack of symmetries in the effects should be
searched for in the causes''. Of course, the occurrence of 
``natural symmetry breaking'' would require a better formulation of
Curie's principle. Symmetries as invariance with respect to a
selected group of transformations are dealt with in physics as 
invariance principles. Explicitly we find the requirement of invariance
in special relativity with respect to the Poincar\'e group (see below),
this brings in the notion of reference system, their equivalence or
lack of. 

The use of transformations for classification purposes has introduced
in physics the transformation method, i.e. a given system is
transformed into an equivalent one considered as a model system
or a ``normal form'' of it. In heuristic terms one may say that to
similar problems there correspond similar solutions (similar means
{\it connected by symmetries}). In analytic mechanics, this approach
has brought in the use of canonical transformations, Hamilton and 
Jacobi. In this approach, evolution itself, the dynamics, is
presented as a 1-parameter group of transformations. In modern physics,
the formulation of special relativity and general relativity on the one
hand, and the extension of canonical transformations in the quantum
setting due to P. Dirac on the other hand, has intimately connected
physics, geometry and transformation groups.  
In mathematics, this research line has culminated in the study of
Lie groups, Lie algebras, representation theory and, in particular,
unitary representations. 

By definition, a {\it symmetry}
of a dynamical system is any transformation which maps bijectively
the set of solutions onto itself. Although not strictly needed,
one can add the further requirement that symmetries should preserve
the parametrization of solutions. In the case of a dynamical system
obeying Newton type equations of motion, i.e.
$$
{dq^{i}\over dt}=u^{i}, \; \; {du^{i}\over dt}=f^{i},
$$
one may or may not require the relation between positions and
velocities to be preserved. In the former case, symmetry 
transformations are further qualified as {\it point symmetries}.

According to the Noether Theorem, if the Lagrangian of a physical
system is invariant under a 1-parameter group of transformations
$\varphi_{\tau}$ on the tangent bundle of configuration space,
then there exists a constant of motion which can immediately be
associated with this invariance, or rather with the 
{\it symmetry group} $\varphi_{\tau}$. It should be remarked that
this association is Lagrangian dependent, i.e. if the equations
of motion admit alternative Lagrangian descriptions, the 
conservation of angular momentum, for instance, may give rise to
different invariance groups.
\vskip 0.3cm
\centerline {\bf 1.2 Simple examples: Euclidean and rotation groups}
\vskip 0.3cm
\noindent
A first important example of symmetry group is here provided by the 
Euclidean group on
${\bf R}^{3}$, i.e. the group of affine transformations which preserve
the length of vectors (the length of vectors being evaluated with
an Euclidean metric $g$). It contains translations and linear homogeneous
transformations which satisfy the condition
$$
g(Tx,Ty)=g(x,y).
\eqno (1.2.1)
$$
By using a basis for ${\bf R}^{3}$, e.g. orthonormal vectors represented
by row or column vectors of the form $(1,0,0),(0,1,0),(0,0,1)$, a 
generic matrix
$$
R \equiv \pmatrix{\alpha_{1} & \alpha_{2} & \alpha_{3} \cr
\beta_{1} & \beta_{2} & \beta_{3} \cr 
\gamma_{1} & \gamma_{2} & \gamma_{3} \cr}
\eqno (1.2.2)
$$
represents a rotation if
$$
g({\vec \alpha},{\vec \alpha})=g({\vec \beta},{\vec \beta})
=g({\vec \gamma},{\vec \gamma})=1,
\eqno (1.2.3)
$$
$$
g({\vec \alpha},{\vec \beta})=g({\vec \alpha},{\vec \gamma})
=g({\vec \beta},{\vec \gamma})=0.
\eqno (1.2.4)
$$
It is easy to show that these $6$ conditions 
imply that $R$ is a rotation matrix if and
only if its transpose equals its inverse. Now we see that
$$
(R_{1}R_{2}) \; { }^{t}(R_{1}R_{2})=R_{1}R_{2} \;
{ }^{t}R_{2} \; { }^{t}R_{1}=\II.
$$
Thus, the product of two rotations is again a rotation, and the
identity matrix is of course a rotation. By virtue of all
these properties, the set of rotations is a subgroup of the group
of linear invertible matrices. The rotation group is denoted
by $O(3)$. It is three-dimensional because one has to subtract
from the dimension of ${\bf R}^{9}$ the number of relations
in Eqs. (1.2.3) and (1.2.4).
Moreover, ${\rm det}R=\pm 1$, and because det is a continuous
function, the matrices with determinant $1$ and those with determinant
$-1$ are two different connected components. 
The set of rotations has only
two components: those expressed by matrices having $-1$ determinant,
which do not form a subgroup since the identity does not belong
to them, and those expressed by matrices having $+1$ determinant.
The latter form the group $SO(3)$ and have the important property
of preserving volumes.
\vskip 10cm
\leftline {\bf References}
\vskip 0.3cm
\item {[1]}
Curie, P. 1894. Sur la symmetrie dans les ph\'enom\`enes physiques.
Symmetrie d'un champ electrique et d'un champ magnetique. 
{\it Journal de Physique}, 3
\item {[2]}
Dirac, P.A.M. 1958. {\it The Principles of Quantum Mechanics},
Oxford: Clarendon Press
\item {[3]}
Marmo, G., Saletan, E.J., Simoni, A. \& Vitale, B. 1983. {\it Dynamical
Systems. A Differential Geometric Approach to Symmetry and 
Reduction}, New York: Wiley
\item {[4]}
Weyl, H. 1952. {\it Symmetry}, Princeton: Princeton University Press
\item {[5]}
Wigner, E.P. 1959. {\it Group Theory and Its Application to the
Quantum Mechanics of Atomic Spectra}, New York: Academic
\vskip 100cm
\centerline {\bf 2. Applications to physical chemistry}
\vskip 0.3cm
\noindent
A very relevant modern application of symmetry concepts in physics is
given by fullerenes. Until 1985, the chemical element Carbon was only
known to exist in two forms, i.e. diamond and graphite. This changed
when Kroto and co-workers discovered an entirely new form of carbon,
which became known as $C_{60}$ or the fullerene molecule. The original
discovery of $C_{60}$ was produced from the laser ablation of graphite.
Since then, other methods of production have been developed. It is
also thought that isolated $C_{60}$ molecules may be found in stars
and interstellar media.

The fullerene molecule (see Figure) 
consists of 60 carbon atoms arranged in
pentagons and hexagons, very like in a standard football. It is
also known as Buckminster Fullerene, by virtue of the resemblance
of this shape to the geodesic domes designed and built by the 
architect R. Buckminster Fuller. More precisely, during experiments
aimed at understanding the mechanism by which long-chain carbon
molecules are formed in interstellar space, graphite was vaporized
by laser irradiation, producing a remarkably stable cluster 
consisting of 60 carbon atoms. To understand what kind of
60-carbon atom structure might give rise to a superstable species,
Kroto and co-authors suggested a truncated icosahedron, a polygon
with 60 vertices and 32 faces, 12 of which are pentagonal and 20
hexagonal. The $C_{60}$ molecule which results when a carbon atom
is placed at each vertex of this structure has all valences satisfied
by two single bonds and one double bond, has many resonance
structures, and appears to be aromatic. In their investigation,
Kroto and co-workers pointed out that, if one considers a tetrahedral
diamond structure, the whole surface of the cluster would be covered
with unsatisfied valences. This led them to look for another plausible
structure which would satisfy all $sp^{2}$ valences. Only a spheroidal
structure appeared likely to satisfy this criterion, and hence they
consulted the studies of Buckminster Fuller. 

\epsfbox{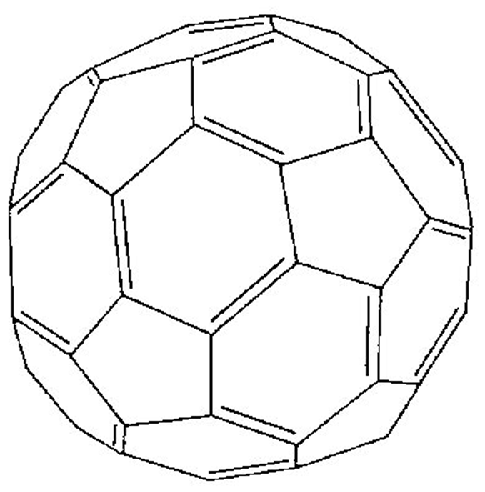}

The Figure shows the 
proposed structure of $C_{60}$ Buckminsterfullerene, the archetype 
of the fullerene family. It has $t$-icosahedral symmetry, as does
the modern European football. One of the properties exploited in
arriving at this model is the fact that a structure in which a
pentagon is completely surrounded by hexagons is stable (Barth and
Lawton 1971). Thus, a cage in which all 12 pentagons are completely
surrounded by hexagons has optimum stability.
\vskip 1cm
\leftline {\bf References}
\vskip 0.3cm
\item {[1]}
Barth, W.E. \& Lawton, R.G. 1971. The synthesis of corannulene. 
{\it J. Am. Chem. Soc.}, 93: 1730-45
\item {[2]}
Kroto, H.W., Heath, J.R., O'Brien, S.C., Curl, R.F. \&
Smalley, R.E. 1985. $C_{60}$: Buckminsterfullerene.
{\it Nature}, 318: 162-63
\item {[3]}
Kroto, H.W. 1987. The stability of the fullerenes $C_{n}$,
with $n=24,28,32,36,50,60$ and $70$. {\it Nature}, 329: 529-31
\item {[4]}
Kroto, H.W. 1997. Symmetry, space, stars and $C_{60}$. 
{\it Rev. Mod. Phys.}, 69: 703-22
\item {[5]}
Smalley, R.E. 1997. Discovering the fullerenes. 
{\it Rev. Mod. Phys.}, 69: 723-30
\vskip 100cm
\centerline {\bf 3. More general groups and related topics}
\vskip 0.3cm
\noindent
Now we move on to a description of more general groups, which
find a wide range of applications in modern theoretical physics.
\vskip 0.3cm
\centerline {\bf 3.1 Galilei group}
\vskip 0.3cm
\noindent
The Galilei group expresses the geometric invariance properties
of the equations of motion of a non-relativistic classical
dynamical system when the system is isolated from external
influences. The general Galilei transformation
$G(R,{\vec v},{\vec \xi},\tau)$ takes a 
point of space-time with coordinates
$x_{1},x_{2},x_{3},t$ to another point with coordinates
$x_{1}',x_{2}',x_{3}',t'$ given by
$$
{\vec x}'=R{\vec x}+{\vec v}t+{\vec \xi},
\eqno (3.1.1)
$$
$$
t'=t+\tau,
\eqno (3.1.2)
$$
where $R \in SO(3)$, ${\vec \xi}$ and ${\vec v}$ 
are fixed vectors in ${\bf R}^{3}$,
and $\tau$ is a real constant. The resulting group multiplication
law is
$$ \eqalignno{
\; & G(R_{2},{\vec v}_{2},{\vec \xi}_{2},\tau_{2})
G(R_{1},{\vec v}_{1},{\vec \xi}_{1},\tau_{1}) \cr
&=G(R_{2}R_{1},R_{2}{\vec v}_{1}+{\vec v}_{2},R_{2}{\vec \xi}_{1}
+{\vec \xi}_{2}+{\vec v}_{2}\tau_{1},
\tau_{2}+\tau_{1}).
&(3.1.3)\cr}
$$
The 1-parameter subgroups are rotations about a fixed axis,
transformations to frames moving in a fixed direction, 
displacements of the origin in a fixed direction, and time
displacements. These correspond to fixed axes of rotation for
$R$ (3 subgroups), fixed directions of ${\vec v}$ (3 subgroups),
fixed directions of ${\vec \xi}$ (3 subgroups) and time displacements
(1 subgroup). The Galilei group is hence 10-dimensional. 
The commutation relations for its Lie algebra are
$$
\left \{ M_{\alpha},M_{\beta} \right \}
=\varepsilon_{\alpha \beta \gamma}M_{\gamma},
\eqno (3.1.4)
$$
$$
\left \{ M_{\alpha},P_{\beta} \right \}
=\varepsilon_{\alpha \beta \gamma}P_{\gamma},
\eqno (3.1.5)
$$
$$
\left \{ M_{\alpha},G_{\beta} \right \}
=\varepsilon_{\alpha \beta \gamma}G_{\gamma},
\eqno (3.1.6)
$$
$$
\left \{ H,G_{\alpha} \right \}=-P_{\alpha},
\eqno (3.1.7)
$$
$$
\left \{ M_{\alpha},H \right \}=\left \{ P_{\alpha},G_{\beta}\right \}
=\left \{G_{\alpha},G_{\beta}\right \}
=\left\{P_{\alpha},P_{\beta}\right\}=\left \{P_{\alpha},H\right \}=0,
\eqno (3.1.8)
$$
where the $M_{\alpha}$ is the infinitesimal generator of rotations
about the $\alpha$ axis, $G_{\alpha}$ of Galilei transformations to
frames moving in the fixed $\alpha$-direction, $P_{\alpha}$ of
displacements of the origin in the $\alpha$-direction, and $H$
of time displacements.
\vskip 100cm
\centerline {\bf 3.2 Lorentz and Poincar\'e groups}
\vskip 0.3cm
\noindent
The Lorentz group is defined 
with the Euclidean metric $g$ being replaced by the Minkowski metric
$\eta$ in Eq. (1.2.1). 
One still has $\eta(Tx,Ty)=\eta(x,y)$, and by virtue of the
signature of the metric, if one writes
$$
T \equiv \pmatrix{\alpha_{0} & \alpha_{1} & \alpha_{2} &
\alpha_{3} \cr
\beta_{0} & \beta_{1} & \beta_{2} & \beta_{3} \cr
\gamma_{0} & \gamma_{1} & \gamma_{2} & \gamma_{3} \cr
\delta_{0} & \delta_{1} & \delta_{2} & \delta_{3} \cr}
\eqno (3.2.1)
$$
one finds
$$
\eta(\alpha,\alpha)=-1, \;
\eta(\beta,\beta)=\eta(\gamma,\gamma)=\eta(\delta,\delta)=1,
\eqno (3.2.2)
$$
$$
\eta(\rho,\sigma)=0 \; \forall \rho \not = \sigma,
\eqno (3.2.3)
$$
for all $\rho,\sigma=\alpha,\beta,\gamma,\delta$. These $10$ independent
conditions determine a $6$-parameter group in the $16$-dimensional
space of $4 \times 4$ matrices. On adding to this 
the linear non-homogeneous
transformations known as space-time translations one gets the
Poincar\'e group. More precisely, this is the
abstract group isomorphic to the geometric group of transformations of
a `world point', and it can be defined independently of any split
of the space-time manifold.

In the applications, one is interested in the following 
realization of the group in terms of transformations
of a world point:
\vskip 0.3cm
\noindent
(i) Space-time displacements: ${\vec r}'={\vec r}+{\vec a}, \; t'=t$ or
${\vec r}'={\vec r}, \; t'=t+b$.
\vskip 0.3cm
\noindent
(ii) Moving frames: 
$$
{\vec r}'={({\vec v} \wedge {\vec r}) \wedge {\vec v} \over v^{2}}
+{{\vec v}\over v^{2}}{{\vec v}\cdot {\vec r}-v^{2}t \over 
\sqrt{1-v^{2}}}, \; |{\vec v}|<1,
\eqno (3.2.4)
$$
$$
t'={{t-{\vec v} \cdot {\vec r}}\over \sqrt{1-v^{2}}}.
\eqno (3.2.5)
$$
(iii) Space rotations: ${\vec r}'=R \; {\vec r}, \; t'=t$.

If we represent the general transformation by $T({\vec a},b,{\vec v},R)$,
with the convention
$$
T({\vec a},b,{\vec v},R)=T({\vec a},0,{\vec 0},\II)
T({\vec 0},b,{\vec 0},\II)
T({\vec 0},0,{\vec v},\II)
T({\vec 0},0,{\vec 0},R),
\eqno (3.2.6)
$$
the effect of a general transformation on $({\vec r},t)$ is
$$
{\vec r}'={\vec a}+{({\vec v} \wedge R {\vec r})\wedge {\vec v} \over v^{2}}
+{{\vec v}\over v^{2}}
{{{\vec v}\cdot R {\vec r}-v^{2}t}\over \sqrt{1-v^{2}}},
\eqno (3.2.7)
$$
$$
t'=b+{{t-{\vec v} \cdot R {\vec r}}\over \sqrt{1-v^{2}}}.
\eqno (3.2.8)
$$
Actually the group that we have so far defined is the connected Lie
sub-group $G_{c}$,
since we have not considered the inversion operations
$$
R_{p}: \; {\vec r}'=-{\vec r}, \; t'=t,
$$
$$
R_{T}: \; {\vec r}'={\vec r}, \; t'=-t,
$$
$$
R_{T}R_{p}=R_{S}: \; {\vec r}'=-{\vec r}, \; t'=-t.
$$
The full group quotiented by the invariant sub-group 
$G_{c}$, connected component containing
the identity, yields a
factor group described by $\II, P, T, TP$. To implement this quotient
group as a sub-group of the original one, 
it is mandatory to introduce, at this stage, a
split of the space-time manifold (see Eqs. (3.2.4)--(3.2.8)),
and its realization depends on the split into space and time.

One can exhibit an inverse for every element $T({\vec a},b,{\vec v},R)$
and show that a multiplication exists which is associative so that one
has a group; moreover the composition functions $z_{r} \equiv f_{r}(x,y)$,
with $x=({\vec a},b,{\vec v},R)$ etc. are differentiable functions of $x$ 
and $y$, so that the system so defined is a Lie group with $10$ parameters,
which is clearly non-commutative. It is customary
to exhibit the $10$ generators in the Lie algebra  
as functions with the help of
a Poisson bracket on $T^{*}{\bf R}^{4}$, the cotangent bundle
of ${\bf R}^{4}$, as:
$$
{\vec p}=\left \{ p_{j} \right \}, \; j=1,2,3; \;
{\rm space} \; {\rm displacement},
$$
$$
H=P_{0}, \; {\rm time} \; {\rm displacement},
$$
$$
{\vec K}=\left \{ K_{j} \right \}, \; j=1,2,3; \;
{\rm moving} \; {\rm frames},
$$
$$
{\vec J}=\left \{ J_{j} \right \}, \; j=1,2,3; \;
{\rm space} \; {\rm rotations}.
$$
The composition law for the Poincar\'e group is then equivalent to the
following bracket relations for the generators:
$$
\left \{P_{j},P_{k}\right \}=0, \;
\left \{P_{j},H \right \}=0, \;
\left \{K_{j},P_{k}\right \}=\delta_{jk}H, \;
\left \{ J_{j},P_{k} \right \}=\varepsilon_{jkl}P_{l},
\eqno (3.2.9)
$$
$$
\left \{ K_{j},H \right \}=P_{j}, \; 
\left \{ J_{j},H \right \}=0,
\eqno (3.2.10)
$$
$$
\left \{ K_{j},K_{k} \right \}=-\varepsilon_{jkl}J_{l}, \;
\left \{ J_{j},K_{k} \right \} =\varepsilon_{jkl}K_{l},
\eqno (3.2.11)
$$
$$
\left \{ J_{j},J_{k} \right \}=\varepsilon_{jkl}J_{l}.
\eqno (3.2.12)
$$

Experience with the rotation, Euclidean and Galilei groups tells
us that, as applied to dynamical systems, generally the infinitesimal
generators (i.e. the elements of the Lie algebra) have an immediate
physical interpretation. We then expect a similar situation to arise for
the Poincar\'e group also. By the close correspondence with the Galilei group,
we identify ${\vec P}$ with linear momentum, $H$ with the energy and
${\vec J}$ with the angular momentum. We may also consider ${\vec K}$ to
be a relativistic `moment'. Bearing this in mind we may now look at the
Poisson-bracket relations; the generators now 
play a dual role. On the one hand
they represent generators of infinitesimal transformations; on the other hand
they represent physical quantities. Thus, the Poisson-bracket relations
$$
\left \{ J_{j},P_{k} \right \}=\varepsilon_{jkl}P_{l}
$$
may be interpreted either as stating that the linear momentum 
transforms like a vector under a rotation (i.e. the increment in ${\vec P}$
in an infinitesimal rotation is at right angles to itself and the axis
of rotation); or as stating that the angular momentum increases by a 
quantity proportional to the normal component of momentum under a
displacement. Similarly, the first of the relations (3.2.10) may be taken
to mean that the energy changes on transforming to a moving frame by a
quantity proportional to the component of linear momentum along the direction
of relative velocity. Equally well it may be taken to mean that the
relativistic moment ${\vec K}$ is not a constant but changes linearly with
respect to time by a quantity proportional to the linear momentum.
\vskip 0.3cm
\centerline {\bf Poincar\'e group and gauge transformations}
\vskip 0.3cm
\noindent
In the investigation of Maxwell's electrodynamics on ${\bf R}^{3}$,
we are familiar with the decomposition of the vector potential $A$ 
into its longitudinal part $A^{L}$ and transverse part $A^{T}$
(this is a corollary of the Kodaira decomposition theorem of the
space of differential forms, bearing in mind that no harmonic
forms exist on a contractible manifold),
such that $dA^{L}=0$ and $\delta A^{T}=0$, $d$ and $\delta$ denoting
exterior differentiation and co-differentiation,
respectively. One may then impose a
supplementary condition, more frequently called gauge choice,
e.g. Lorenz or Coulomb or axial or temporal,
to achieve a particular form of the field equations expressed in
terms of the potential (and also to ensure that the operator
occurring in the classical theory of small disturbances is
invertible). At that stage, all calculations rely on a gauge choice
made in a given reference frame. However, to obtain a
Poincar\'e-covariant formulation one needs a transformation
that takes us from 
$$
\delta A^{T}=0
\eqno (3.2.13)
$$
in a given frame, to the same form
$$
\delta' {A'}^{T}=0
\eqno (3.2.14)
$$
in a different frame. For this to be the case, it is necessary
{\it to change the action of the Poincar\'e group on the vector
potential}. Of course, we can transform the vector potential by 
adding to it a closed 1-form so that the resulting electromagnetic
field is not affected. 

If $X$ is any vector field in the Poincar\'e algebra, one can
associate with it an operator $D_{X}$ the action of which is
given by
$$
D_{X}A=L_{X}A+df_{X}.
\eqno (3.2.15)
$$
It can be proved that the appropriate functions can be so chosen
that $D_{X}$ provides a representation of the Poincar\'e algebra
$$
D_{X}D_{Y}-D_{Y}D_{X}=D_{[X,Y]}.
\eqno (3.2.16)
$$
This is called a {\it gauge-dependent representation of the
Poincar\'e algebra}, which maps Eq. (3.2.13) into Eq. (3.2.14).
\vskip 1cm
\leftline {\bf References}
\vskip 0.3cm
\item {[1]}
Sudarshan, E.C.G. \& Mukunda, N. 1974. {\it Classical Dynamics:
A Modern Perspective}, New York: Wiley.
\item {[2]}
DeWitt, B.S. 2003. {\it The Global Approach to Quantum
Field Theory}, Oxford: Oxford University Press.
\item {[3]}
Barut, A.O. 1980. {\it Electrodynamics and Classical Theory of
Fields and Particles}, New York: Dover.
\item {[4]}
Marmo, G. 1986. {\it Differential Forms and Electrodynamics}
(Lectures at Shanxi University).
\vskip 100cm
\centerline {\bf 3.3 Conformal Transformations}
\vskip 0.3cm
\noindent
Given the space-time manifold $(M,g)$, a {\it conformal rescaling} of
the metric $g=g_{ab}dx^{a} \otimes dx^{b}$ is a map
$$
g_{ab} \rightarrow {\widehat g}_{ab}=\Omega^{2} g_{ab},
\eqno (3.3.1)
$$
where $\Omega$ is a smooth and positive scalar function on $M$.
If $g_{ab}$ is the Minkowski metric 
$\eta_{ab}={\rm diag}(-1,1,1,1)$, the new metric 
${\widehat \eta}_{ab}$ is not flat in general, but there are
exceptions: the Riemann tensor obtained from
${\widehat \eta}_{ab}$ when $\Omega$ is constant or equal
to $(x^{a}x_{a})^{-1}$ does indeed vanish. 

If $(M_{1},g)$ and $(M_{2},h)$ 
are two pseudo-Riemannian manifolds, then a diffeomorphism
$$
f: (M_{1},g) \rightarrow (M_{2},h)
$$
is said to be a {\it conformal mapping} if the metric induced 
on $M_{2}$ by $f$ is a conformal
rescaling of the given metric on $M_{2}$. Two standard but
important examples are as follows.
\vskip 0.3cm
\noindent
(i) If both $M_{1}$ and $M_{2}$ are Minkowski space-time,
and $f$ is the map
$$
f: x^{a} \rightarrow k x^{a},
\eqno (3.3.2)
$$
with $k$ a positive constant, then the induced metric
${\widehat \eta}_{ab}$ is related to the original metric by
${\widehat \eta}_{ab}=k^{-2}\eta_{ab}$. Such a conformal 
mapping is called a {\it dilation}.
\vskip 0.3cm
\noindent
(ii) Let $M_{1}$ be Minkowski space-time minus the null cone
of $p^{a}$, while $M_{2}$ is chosen to be Minkowski space-time
minus the null cone of the origin. Define the map $f$ by
$$
f: x^{a} \rightarrow {p^{a}-x^{a} \over (p_{b}-x_{b})
(p^{b}-x^{b})}.
\eqno (3.3.3)
$$
Such a map is a diffeomorphism, and the induced metric 
${\widehat \eta}_{ab}$ on $M_{2}$ is related to the original
metric $\eta_{ab}$ on $M_{2}$ by 
${\widehat \eta}_{ab}=\Omega^{2}\eta_{ab}$, with 
$\Omega(x) \equiv (x^{a}x_{a})^{-1}$. The resulting conformal
mapping is called an {\it inversion}. The next step consists in
adjoining to Minkowski space-time a null cone at infinity, here
denoted by $\cal I$. Hence one gets a compact manifold $M$ containing
Minkowski space-time, and the inversion map $f$ can be extended
to a conformal map from $M$ to itself.

The Lie group of conformal mappings from $M$ to itself is the
{\it conformal group} and is denoted by $C(1,3)$. It is 
15-dimensional in $4$ space-time dimensions, being generated
by the Poincar\'e transformations ($6$ parameters from Lorentz
and $4$ from translations), the dilations ($1$ parameter) and the
inversions ($4$ parameters).

When conformal transformations are considered, two meanings can
be attached to the statement that a field theory is conformally
invariant. A first option is to require that it should be possible
to assign a conformal weight to each field in such a way that 
the resulting field equations are invariant under arbitrary
conformal rescalings of the metric. The fields of the theory
(either scalar, or tensor or spinor) are said to have conformal
weight $k$ if they transform according to 
$\varphi \rightarrow \Omega^{k}\varphi$ under a conformal 
rescaling of the metric. The consideration of Minkowski
space-time is sufficient to show an important property: since under
a dilation the d'Alembert operator $\cstok{\ }$ transforms as
$\cstok{\ } \rightarrow k^{-2} \cstok{\ }$, the field equation
$(\cstok{\ }+m^{2})\varphi=0$ is not invariant unless the mass
parameter vanishes. Thus, to be conformally invariant, a field
theory has to involve only massless fields.

To define conformal invariance one can however require that the
field theory should be invariant under the conformal group 
$C(1,3)$. This is the second possible definition. If a theory
is both Poincar\'e-invariant and invariant under conformal 
rescalings, it is then conformally invariant in this second
sense as well. This holds because the Poincar\'e transformations
become conformal transformations according to any other 
conformally rescaled flat metric, and the resulting conformal
transformations, jointly with Poincar\'e transformations,
generate the whole conformal group.
\vskip 1cm
\leftline {\bf References}
\vskip 0.3cm
\item {[1]}
Penrose, R. 1974. Relativistic symmetry groups, in
{\it Group Theory in Nonlinear Problems}, ed. A.O. Barut,
Dordrecht: Reidel Publishing Company
\item {[2]}
Ward, R.S. \& Wells, R.O. 1990. {\it Twistor Geometry and 
Field Theory}, Cambridge: Cambridge University Press
\vskip 100cm
\centerline {\bf 3.4 Laplace Equation}
\vskip 0.3cm
\noindent
Let us consider functions $u$ depending on $n$ variables 
$x_{1},x_{2},...,x_{n}$ in a domain $G$ of ${\bf R}^{n}$ with
boundary $\Gamma$. The differential equation
$$
\bigtriangleup u=\sum_{l=1}^{n}
{\partial^{2}u \over \partial x_{l}^{2}}=0
\eqno (3.4.1)
$$
is called the Laplace equation, and its solutions are called
{\it harmonic functions}. The harmonic functions belong therefore
to the kernel of the Laplace operator $\bigtriangleup$.

Solutions of the Laplace equation which have continuous second 
derivatives in an open, connected and bounded region $G$ of 
space are called {\it regular} in $G$. In $2$ dimensions the
general solution of the Laplace equation is the real part of
any analytic function of the complex variable $x+iy$. In $3$ 
dimensions it remains easy to construct solutions which depend
on arbitrary functions. For example, let $f(w,t)$ be analytic
in the complex variable $w$ for fixed real $t$. When the variable
$t$ is allowed to take arbitrary values, both the real and 
imaginary parts of the function
$$
u \equiv f(z+ix\cos t +iy \sin t, t)
$$
of the real variables $x,y,z$ are solutions of the Laplace
equation. Further solutions may be obtained by superposition
via an integral representation, i.e.
$$
u=\int_{a}^{b} f(z+ix \cos t+iy \sin t,t)dt.
\eqno (3.4.2)
$$
For example, on setting $f(w,t)=w^{n}e^{iht}$ for some integers
$n$ and $h$, integration from $-\pi$ to $+\pi$ yields the
homogeneous polynomials
$$
u=\int_{-\pi}^{\pi}(z+ix \cos t +iy \sin t)^{n}e^{iht}dt
=r^{n}e^{ih\phi}P_{n,h}(\cos \theta),
\eqno (3.4.3)
$$
having introduced polar coordinates in ${\bf R}^{3}$,
with $P_{n,h}$ the standard notation for Legendre functions.

By transforming to polar coordinates in the plane or in space,
the action of the Laplacian becomes
$$
\bigtriangleup u={1\over r}\left[{\partial \over \partial r}
(ru_{r})+{\partial \over \partial \phi}\left({u_{\phi}\over r}
\right)\right],
\eqno (3.4.4)
$$
$$
\bigtriangleup u={1\over r^{2}\sin \theta}\left[
{\partial \over \partial r}(r^{2}u_{r}\sin \theta)
+{\partial \over \partial \theta}(u_{\theta}\sin \theta)
+{\partial \over \partial \phi}
\left({u_{\phi}\over \sin \theta}\right)\right],
\eqno (3.4.5)
$$
respectively. These formulae, and their extension to ${\bf R}^{n}$,
make it possible to prove the following theorem: {\it if 
$u(x_{1},...,x_{n})$ is a regular harmonic function in the domain
$G$, the function
$$
v(x_{1},...,x_{n})=r^{-(n-2)}u \left({x_{1}\over r^{2}},...,
{x_{n}\over r^{2}}\right)
\eqno (3.4.6)
$$
also satisfies the Laplace equation and is regular in the region
$G'$ obtained from $G$ by inversion with respect to the unit
sphere}. We therefore learn that, apart from the factor 
$r^{2-n}$, the harmonic character of a function is invariant
under inversions with respect to spheres. Moreover, the harmonic
property is retained completely under similarity transformations,
translations, and simple reflections across planes.

Let the function $u$ be regular and harmonic in a bounded domain
$G$. If we invert $G$ with respect to a sphere of unit radius
whose centre lies in $G$, the interior of $G$ is carried into the
exterior $G'$ of the inverted boundary surface $\Gamma'$. The
harmonic function in Eq. (3.4.6) is then called regular in this exterior
region $G'$. Regularity in a domain $G$ extending to infinity is
therefore defined as follows: the domain $G$ is inverted with
respect to a sphere with centre outside of $G$, so that $G$ is
transformed into a bounded domain $G'$. By definition, the
harmonic function $u$ is called regular in $G$ if the above function
$v$ is regular in $G'$. In particular, $u$ is said to be {\it regular
at infinity} if $G$ contains a neighbourhood of the point at infinity
and a value is assigned to the function $u$ at the point at infinity
such that $v$ is regular in the bounded domain $G'$. These definitions
imply that the constant function is regular at infinity in the plane,
but not in spaces of $3$ or more dimensions. In space, for arbitrary
values of $a$, the functions
$$
u \equiv 1-a+{a\over r}
$$
are harmonic outside the unit sphere and take the boundary value
$u=1$ on the sphere. But $u={1\over r}$ is the only function of
this family which is regular in the region exterior to the
unit sphere. 

For any number $n$ of dimensions, the only solutions of the Laplace
equation which depend only on the distance $r$ of the point $x$
from a fixed point $\xi$, are (up to arbitrary multiplicative
and additive constants) the functions ($\omega_{n}$ being the
surface area of the unit sphere $S^{n}$)
$$
\gamma(r)={1\over (n-2)\omega_{n}}r^{2-n}, 
\; \; {\rm if} \; \; n>2, 
\eqno (3.4.7)
$$
$$
\gamma(r)={1\over 2\pi}\log {1\over r},
\; \; {\rm if} \; \; n=2.
\eqno (3.4.8)
$$
These exhibit the so-called characteristic singularity at $r=0$.
Every solution of the Laplace equation in the domain $G$ of
the form (here $r^{2} \equiv \sum_{l=1}^{n}(x_{l}-\xi_{l})^{2}$)
$$
\psi(x_{1},...,x_{n};\xi_{1},...,\xi_{n})=\gamma(r)+w,
\eqno (3.4.9)
$$
for $\xi$ inside $G$ and $w$ regular, is said to be a 
{\it fundamental solution} with a singularity at the parameter
point $\xi$.

Rotational invariance of the Laplace equation (3.4.1) is proved in
an elegant way by pointing out that the symbol $\sigma$ (or characteristic
polynomial) of the Laplace operator therein reads
$$
\sigma(k)=k_{x_{1}}^{2}+...+k_{x_{n}}^{2},
\eqno (3.4.10)
$$
which is the squared length of the $n$-component vector
$\vec k$ in momentum space, and it is well known that rotations
preserve the length of vectors. Note that, when writing Eq. (3.4.10),
we end up defining the Laplace operator with a minus sign in
front of all second derivatives. This has two related advantages:
(i) it yields a Laplace operator with a spectrum bounded from
below on compact Riemannian manifolds; (ii) it leads to a symbol
which is a positive-definite quadratic form.

Operators which differ from the Laplacian by the addition of a
potential term are said to be of Laplace type. In modern field
theories of fundamental interactions, these operators act on
sections of vector bundles $V$ over a base space which is a
Riemannian manifold $(M,g)$, and read as
$$
P \equiv -g^{ab}\nabla_{a}\nabla_{b}-E,
\eqno (3.4.11)
$$
where $\nabla$ is the connection on $V$, and $E$ is an 
endomorphism of $V$ (i.e. the potential term in the operator
of Laplace type). Equation (3.4.11) expresses the general form
of the operator acting on the gauge field whenever linear covariant
gauges are used for (path-integral) quantization.
\vskip 1cm
\leftline {\bf References}
\vskip 0.3cm
\item {[1]}
Kellogg, O.D. 1954. {\it Foundations of Potential Theory},
New York: Dover
\item {[2]}
Courant, R. \& Hilbert, D. 1961. {\it Methods of Mathematical
Physics. II. Partial Differential Equations}, New York:
Interscience
\item {[3]}
Garabedian, P.R. 1964. {\it Partial Differential Equations},
New York: Chelsea
\item {[4]}
Avramidi, I.G. \& Esposito, G. 1999. Gauge theories on manifolds
with boundary. {\it Communications in Mathematical Physics},
200: 495-543
\vskip 100cm
\noindent
\centerline {\bf 3.5 Quantum Groups}
\vskip 0.3cm
\noindent
Several current developments in group and field theory
rely on the concept of quantum group. Such a name is an
abuse of language, since one actually deals with a Hopf 
algebra consisting of the following structures:
\vskip 0.3cm
\noindent
(i) A unital algebra $H$ over the complex field $\cal C$ (i.e. a
linear space $H$ endowed with multiplication,
say $m: H \times H \rightarrow {\cal C}$, and unity maps which
are complex-linear, satisfy the associativity of multiplication
and the existence of unit element);
\vskip 0.3cm
\noindent
(ii) A coproduct $\bigtriangleup: H \rightarrow H \otimes H$
and counit $\varepsilon: H \rightarrow {\cal C}$ forming a 
coalgebra, with $\bigtriangleup, \varepsilon$ algebra homomorphisms;
\vskip 0.3cm
\noindent
(iii) An antipode $S: H \rightarrow H$ such that 
$(S \otimes {\rm id})\bigtriangleup =i \varepsilon
=({\rm id} \otimes S)\bigtriangleup$.

A coalgebra is just like an algebra but with the arrows
on the maps occurring in the axioms reversed. Thus, the
coassociativity and counity axioms are
$$
(\bigtriangleup \otimes {\rm id})\bigtriangleup
=({\rm id} \otimes \bigtriangleup)\bigtriangleup,
\eqno (3.5.1)
$$
$$
(\varepsilon \otimes {\rm id})\bigtriangleup
=({\rm id} \otimes \varepsilon)\bigtriangleup={\rm id}.
\eqno (3.5.2)
$$
The antipode plays a role that generalizes the concept of group
inversion. Other than that the only new mathematical structure
that the reader has to contend with is the coproduct 
$\bigtriangleup$ and its associated counit. There are several
ways of interpreting the meaning of this, depending on our point
of view. If the quantum group is like the enveloping algebra 
$U(g)$ generated by a Lie algebra $g$, one should think of 
$\bigtriangleup$ as providing the rule by which actions extend to
tensor products. Thus, $U(g)$ is trivially a Hopf algebra with
$$
\bigtriangleup \xi=\xi \otimes 1 + 1 \otimes \xi, \; \;
\forall \xi \in g,
\eqno (3.5.3)
$$
which says that when a Lie algebra element $\xi$ acts on tensor
products it does so by $\xi$ in the first factor and then $\xi$
in the second factor. Similarly it says that when a Lie algebra
acts on an algebra it does so as a derivation. On the other hand,
if the quantum group is like a coordinate algebra ${\cal C}[G]$, 
then $\bigtriangleup$ expresses the group multiplication and
$\varepsilon$ the group identity element ${\rm e}$. Thus, if 
$f \in {\cal C}[G]$, the coalgebra is
$$
(\bigtriangleup f)(g,h)=f(gh), \; \;
\forall g,h \in G, \; \;
\varepsilon f =f({\rm e}),
\eqno (3.5.4)
$$
at least for suitable choices of $f$. In other words, it 
expresses the group product $G \times G \rightarrow G$ by a map
in the reversed direction in terms of coordinate algebras. 

{}From yet another point of view $\bigtriangleup$ simply makes the
dual $H^{*}$ also into an algebra. Hence a Hopf algebra is basically
and algebra such that the dual $H^{*}$ is also an algebra, in a
compatible way. For every finite-dimensional $H$ there is a dual
$H^{*}$, and similarly in the infinite-dimensional case,
where one has to distinguish between algebraic and topological duals.
Among the many examples that can be given, we here mention only
two, for length reasons.
\vskip 0.3cm
\noindent
(i) The Planck scale quantum group generated by position $x$ and
momentum $p$ with commutation relations (here $l$ is a parameter
having dimension length) 
$$
[x,p]=i{\hbar}\Bigr(1-e^{-x/l}\Bigr),
\eqno (3.5.5)
$$
and coproduct
$$
\bigtriangleup x=x \otimes 1 + 1 \otimes x,
\eqno (3.5.6)
$$
$$
\bigtriangleup p=p \otimes e^{-x/l}+1 \otimes p.
\eqno (3.5.7)
$$
\vskip 0.3cm
\noindent
(ii) Quantum groups may be viewed as deformed enveloping algebras.
The simplest example is the quantum group $U_{q}(su_{2})$ with
generators $H,X_{\pm}$ and defining relations
$$
[H,X_{\pm}]=\pm 2 X_{\pm},
\eqno (3.5.8)
$$
$$
[X_{+},X_{-}]={q^{H}-q^{-H} \over q-q^{-1}},
\eqno (3.5.9)
$$
and coproduct
$$
\bigtriangleup X_{\pm}=X_{\pm} \otimes q^{H/2}
+q^{-H/2} \otimes X_{\pm},
\eqno (3.5.10)
$$
$$
\bigtriangleup H = H \otimes 1 + 1 \otimes H.
\eqno (3.5.11)
$$
The coproduct $\bigtriangleup$ here is a deformation of the
usual additional one, which is recovered as $q \rightarrow 1$.
The deformation modifies how an action of $X_{\pm}$ extends
to tensor products.

More recently, groupoids and algebroids have been introduced,
and these concepts have enlarged the notion of symmetry. Roughly
speaking, it is like going from the family of transformations of
an infinite tiling to the transformation group of a finite tiling.
\vskip 10cm
\leftline {\bf References}
\vskip 0.3cm
\item {[1]}
Majid, S. 1995. {\it Foundations of Quantum Group Theory},
Cambridge: Cambridge University Press
\item {[2]}
Majid, S. 2000. Quantum groups and concommutative geometry.
{\it Journal of Mathematical Physics}, 41: 3892-942
\item {[3]}
Gracia-Bondia, J.M., Lizzi, F., Marmo, G. \& Vitale, P. 2002.
Infinitely many star products to play with. 
{\it JHEP}, 0204:026
\item {[4]}
Majid, S. 2002. {\it A Quantum Groups Primer}. London Mathematical
Society Lecture Note Series, 292
\vskip 100cm
\centerline {\bf 3.6 Sklyanin Algebra}
\vskip 0.3cm
\noindent
One of the most powerful methods for studying the exactly solvable
models of quantum and statistical physics is the quantum inverse
problem method. The problem of {\it enumerating} the discrete quantum
systems which can be solved by the quantum inverse problem method
reduces to the problem of enumerating the operator-valued
functions $L(u)$ that satisfy the equation
(here $L' \equiv L \otimes 1$ and $L'' \equiv 1 \otimes L$)
$$
R(u-v)L'(u)L''(v)=L''(v)L'(u)R(u-v)
\eqno (3.6.1)
$$
for a fixed solution $R(u)$ of the quantum Yang--Baxter equation
$$
R_{12}(u-v)R_{13}(u)R_{23}(v)=R_{23}(v)R_{13}(u)R_{12}(u-v).
\eqno (3.6.2)
$$
In the classical theory, Eq. (3.6.1) is replaced by
(here curly brackets with a comma $\left \{ \; , \; \right \}$ denote
the Poisson bracket, and $[A,B]_{\pm} \equiv AB \pm BA$ for 
given matrices $A$ and $B$)
$$
\left \{ L'(u),L''(v) \right \}
=[r(u-v),L'(u)L''(v)]_{-},
\eqno (3.6.3)
$$
while Eq. (3.6.2) becomes the classical Yang--Baxter equation
$$
[r_{12}(u-v),r_{13}(u)]_{-}+[r_{12}(u-v),r_{23}(v)]_{-}
+[r_{13}(u),r_{23}(v)]_{-}=0.
\eqno (3.6.4)
$$
Equations (3.6.1) and (3.6.3) provide a systematic procedure for obtaining
completely integrable lattice approximations to various 
continuous completely integrable systems. We here summarize the 
classical and quantum investigations of Sklyanin, with emphasis
on the associated algebras.
\vskip 0.3cm
\noindent
(i) {\it Classical Theory}. Let $r(u)$ be the simplest solution of
Eq. (3.6.4):
$$
r(u)=\sum_{\alpha=1}^{3}w_{\alpha}(u)
\sigma_{\alpha}\otimes \sigma_{\alpha},
\eqno (3.6.5)
$$
where $\sigma_{\alpha}$ are the Pauli matrices and the coefficients 
$w_{\alpha}(u)$ can be expressed in terms of the Jacobi 
elliptic functions as
$$
w_{1}(u)=\rho {1\over {\rm sn}(u,k)}, \;
w_{2}(u)=\rho {{\rm dn}(u,k)\over {\rm sn}(u,k)}, \;
w_{3}(u)=\rho {{\rm cn}(u,k)\over {\rm sn}(u,k)},
\eqno (3.6.6)
$$
with fixed values of $\rho >0$ and $k \in [0,1]$. On looking for
solutions of Eq. (3.6.3) in the form
$$
L(u)=S_{0}+i \sum_{\alpha=1}^{3}w_{\alpha}(u)
S_{\alpha}\sigma_{\alpha},
\eqno (3.6.7)
$$
one eventually finds the following quadratic algebra of Poisson
brackets for the variables $S_{\alpha}$:
$$
\left \{ S_{\alpha},S_{0} \right \}
=2J_{\beta \gamma}S_{\beta}S_{\gamma}, 
\eqno (3.6.8)
$$
where the right-hand side carries indices $\beta,\gamma$ summed over 
and ranging from $1$ to $3$, while
$$ 
\left \{ S_{\alpha},S_{\beta} \right \}=-2S_{0}S_{\gamma},
\eqno (3.6.9)
$$
where $\alpha \not = \beta \not = \gamma$.
The constants $J_{\alpha \beta}$ make it possible to express
that the coefficients $w_{\alpha}$ lie on a quadric, i.e.
$$
w_{\alpha}^{2}-w_{\beta}^{2}=J_{\alpha \beta}.
\eqno (3.6.10)
$$
Equations (3.6.8) and (3.6.9) define a quadratic homogeneous Poisson
brackets Lie algebra, here denoted by $\cal P$. Such equations can be
presented in a neater and generalized 
way upon remarking that, given the differential
form (here $i,j=0,1,2,3$)
$$
F=d(a_{i}x_{i}^{2})\wedge d(b_{j}x_{j}^{2})
=2(a_{i}b_{j}-b_{i}a_{j})x_{i}x_{j}dx_{i}\wedge dx_{j},
\eqno (3.6.11)
$$
the contraction of $F$ with a volume element expressed in 
contravariant form defines a Poisson tensor $\Lambda$ such
that $\Lambda^{lm}=\varepsilon^{lmjk}F_{jk}$. The tensor
$\Lambda$ defines therefore the following Poisson structure 
on ${\bf R}^{4}$:
$$
\Lambda=\varepsilon_{ijkl}(a_{i}b_{j}-b_{i}a_{j})
x_{i}x_{j}\partial_{k} \wedge \partial_{l},
\eqno (3.6.12)
$$
and hence we get the Poisson bracket
$$
\left \{ x_{k},x_{l} \right \} = \varepsilon_{klij}
(a_{i}b_{j}-b_{i}a_{j})x_{i}x_{j}.
\eqno (3.6.13)
$$
In particular, on setting $b_{0}=0,b_{1}=b_{2}=b_{3}=1,a_{0}=1$,
we get the Sklyanin bracket
$$
\left \{ x_{k},x_{l} \right \}=\varepsilon_{jkl}x_{0}x_{j}, \; \; \; \; 
\left \{ x_{k},x_{0} \right \} = \varepsilon_{jkl}
(a_{j}-a_{l})x_{j}x_{l},
\eqno (3.6.14)
$$
where now $j,k,l=1,2,3$.
\vskip 0.3cm
\noindent
(ii) {\it Quantum Theory}. Let $R(u)$ be the solution of Eq. (3.6.2)
given by
$$
R(u)=1+\sum_{\alpha=1}^{3}W_{\alpha}(u)
\sigma_{\alpha} \otimes \sigma_{\alpha},
\eqno (3.6.15)
$$
with coefficients $W_{\alpha}(u)$ lying on the algebraic curve
(it is a standard notation, 
in this context, not to write the index $\gamma$ on
the right-hand side)
$$
{{W_{\alpha}^{2}-W_{\beta}^{2}}\over {W_{\gamma}^{2}-1}}
={\bf J}_{\alpha \beta} 
\eqno (3.6.16)
$$
and expressed in the form
$$
W_{1}(u)={{\rm sn}(i\eta,k) \over {\rm sn}(u+i \eta,k)},
\eqno (3.6.17)
$$
$$
W_{2}(u)={{\rm dn}\over {\rm sn}}(u+i\eta,k)
{{\rm sn}\over {\rm dn}}(i\eta,k),
\eqno (3.6.18)
$$
$$
W_{3}(u)={{\rm cn}\over {\rm sn}}(u+i\eta,k)
{{\rm sn}\over {\rm cn}}(i\eta,k).
\eqno (3.6.19)
$$
In analogy with the classical theory, one looks for the solution
${\bf L}(u)$ of Eq. (3.6.1) in the form
$$
{\bf L}(u)={\bf S}_{0}+\sum_{\alpha=1}^{3}W_{\alpha}(u)
{\bf S}_{\alpha},
\eqno (3.6.20)
$$
where the variables ${\bf S}_{\alpha}$ are found to obey the 
commutation relations
$$
\Bigr[{\bf S}_{\alpha},{\bf S}_{0}\Bigr]_{-}
=-i {\bf J}_{\beta \gamma}  
\Bigr[{\bf S}_{\beta},{\bf S}_{\gamma}\Bigr]_{+},
\eqno (3.6.21)
$$
$$
\Bigr[{\bf S}_{\alpha},{\bf S}_{\beta}\Bigr]_{-}
=i \Bigr[{\bf S}_{0},{\bf S}_{\gamma}\Bigr]_{+}.
\eqno (3.6.22)
$$
Equations (3.6.21) and (3.6.22) generate a two-sided ideal $I$ in the free
associative algebra $\cal A$ whose four generators are the
variables ${\bf S}_{\alpha}$. 

If we set $\eta=\rho h$ in Eqs. (3.6.17)--(3.6.19) 
and pass to the limit of vanishing Planck constant, we obtain
the limiting relations
$$
W_{\alpha}(u)=ihw_{\alpha}(u)+{\rm O}(h^{2}),
\eqno (3.6.23)
$$
$$
R(u)=1+ihr(u)+{\rm O}(h^{2}),
\eqno (3.6.24)
$$
$$
{\bf J}_{\alpha \beta} 
=h^{2}J_{\alpha \beta}+{\rm O}(h^{4}).
\eqno (3.6.25)
$$
On using the expansions (3.6.23)--(3.6.25) and assuming that the quantum
quantities ${\bf S}_{\alpha}$ become the corresponding classical
quantities $S_{\alpha}$ when $h \rightarrow 0$, according
to the prescription
$$
{\bf S}_{0} \sim h S_{0}, \;
{\bf S}_{\alpha} \sim S_{\alpha},
$$
the equations (3.6.1), (3.6.2) and (3.6.21), 
(3.6.22) for the quantum theory are transformed
into the equations (3.6.3), (3.6.4), and (3.6.8), (3.6.9) for the
classical theory. Moreover, ${\bf L}(u) \sim h L(u)$ and hence
the quadratic Poisson bracket algebra $\cal P$ defined by (3.6.8)
and (3.6.9) is the classical limit of the quotient algebra
${\cal F} \equiv {{\cal A}\over {\cal I}}$.   

Relevant examples of irreducible, finite-dimensional, 
self-adjoint representations of the algebra $\cal F$ are given,
in two dimensions, by the Pauli matrices, i.e.
${\bf S}_{0}=1$ and ${\bf S}_{\alpha}=\sigma_{\alpha}$, while in
three dimensions, on defining the ${\bf J}_{\alpha}$ through
(cf. Eq. (3.6.16))
$$
{\bf J}_{\alpha \beta} 
=-{{{\bf J}_{\alpha}-{\bf J}_{\beta}}\over
{\bf J}_{\gamma}},
\eqno (3.6.26)
$$
one finds
$$
{\bf S}_{0} = \pmatrix{{\bf J}_{3} & 0 & {\bf J}_{1}-{\bf J}_{2} \cr
0 & {\bf J}_{1}+{\bf J}_{2}-{\bf J}_{3} & 0 \cr
{\bf J}_{1}-{\bf J}_{2} & 0 & {\bf J}_{3} \cr},
\eqno (3.6.27)
$$
$$
{\bf S}_{1}=\sqrt{2{\bf J}_{2}{\bf J}_{3}}
\pmatrix{0 & 1 & 0 \cr 1 & 0 & 1 \cr 0 & 1 & 0 \cr},
\eqno (3.6.28)
$$
$$
{\bf S}_{2}=\sqrt{2{\bf J}_{3}{\bf J}_{1}}
\pmatrix{0 & -i & 0 \cr i & 0 & -i \cr 0 & i & 0 \cr},
\eqno (3.6.29)
$$
$$
{\bf S}_{3}=2\sqrt{{\bf J}_{1}{\bf J}_{2}}
\pmatrix{1 & 0 & 0 \cr 0 & 0 & 0 \cr 0 & 0 & -1 \cr}.
\eqno (3.6.30)
$$
The representation expressed by Eqs. (3.6.27)--(3.6.30) is self-adjoint
only when ${\bf J}_{\alpha}>0$.
\vskip 1cm
\leftline {\bf References}
\vskip 0.3cm
\item {[1]}
Sklyanin, E.K. 1982. Some algebraic structures connected with the
Yang--Baxter equation. {\it Functional Analysis and Its Applications},
16: 263-70
\item {[2]}
Quano, Y.H. 1991. Generalized Sklyanin algebra. {\it Modern Physics
Letters}, A6: 3635-40
\item {[3]}
Majid, S. 1993. Braided matrix structure of the Sklyanin algebra
and of the quantum Lorentz group. {\it Communications in
Mathematical Physics}, 156: 607-38
\item {[4]}
Quano, Y.H. 1994. Sklyanin algebra and integrable lattice models.
{\it International Journal of Modern Physics}, A9: 2245-81
\item {[5]}
Grabowski, J., Marmo, G. \& Perelomov, A.M. 
1993. Poisson structures:
towards a classification. {\it Modern Physics Letters}, 
A8: 1719-33 
\vskip 0.3cm
\noindent
{\bf Acknowledgments}. The authors acknowledge partial financial
support by PRIN 2002 {\it SINTESI}.

\bye